\documentclass[pre,showpacs,aps,twocolumn,superscriptaddress]{revtex4}
\usepackage{amssymb,amsmath,amsthm}
\usepackage[dvips]{graphicx}

\newcommand{\vect}[1]{\boldsymbol{#1}}

\newcommand{\mydef}
        {\stackrel{\mathrm{def}}{=}}

\begin{document}
\title{Anomalous diffusion of dissipative solitons in the cubic-quintic complex Ginzburg-Landau equation in two
spatial dimensions}

\date{\today}

\author{Jaime Cisternas}
\email[Corresponding author: ]{jecisternas@miuandes.cl}
\affiliation{Facultad de Ingenier\'{\i}a y Ciencias Aplicadas, Universidad de los Andes, Chile}
\author{Orazio Descalzi}
\affiliation{Facultad de Ingenier\'{\i}a y Ciencias Aplicadas, Universidad de los Andes, Chile}
\author{Tony Albers}
\affiliation{Institute of Physics, Technische Universit\"at Chemnitz, 09107 Chemnitz, Germany}
\author{G\"unter Radons}
\affiliation{Institute of Physics, Technische Universit\"at Chemnitz, 09107 Chemnitz, Germany}

\begin{abstract}
We demonstrate the occurrence of anomalous diffusion of dissipative solitons
in a `simple' and deterministic prototype model: the cubic-quintic
complex Ginzburg-Landau equation in two spatial dimensions. The main features of their dynamics, induced by
symmetric-asymmetric explosions, can be modeled by a subdiffusive continuous-time random walk,
while in the case dominated by only asymmetric explosions it becomes characterized by normal diffusion.
\end{abstract}

 \maketitle

\noindent
Dissipative localized structures, objects whose existence is based on the
delicate balance of nonlinearity, dispersion, gain and loss \cite{nailbook},
have been widely observed experimentally in the context of many
different branches of science \cite{LS,DS}, ranging from nonlinear
optics \cite{weiss,stegeman} to hydrodynamics \cite{kolodner3,ahlers},
including chemical surface reactions \cite{harm1},
reaction-diffusion systems \cite{DS}, granular matter \cite{melo},
colloidal suspensions \cite{fineberg}, etc. Regardless
of their specific field they share some characteristics: they appear in
different geometries; they typically are either stationary, oscillating or
move with fixed shape at constant velocity (except when they interact
with other dissipative structures or suffer an external forcing). Nevertheless, there is
limited evidence of localized structures experiencing random walks
(an interesting exception is the motion of current density filaments \cite{DS}).
Even in the case of spatiotemporal chaotic localized structures, they do not move in space
significantly \cite{marcel}.

\noindent
Random walks that are qualitatively different from normal diffusion have been recently observed
in different contexts. Superdiffusion (with step lengths described by a long-tailed probability distribution)
has been reported for an ensemble of ultracold Rb atoms \cite{sagi} and for optical materials \cite{barthe},
while subdiffusion (with waiting times described by a heavy-tailed distribution)
has been observed in the random motion of
individual mRNA molecules inside {\it E. coli} \cite{GC06}.
Superdiffusion and subdiffusion, in different regimes, can be observed in single particle tracking of
polystyrene microbeads in micellar solutions \cite{jeon}. Transition from
subdiffusive to normal diffusion was observed for telomeres in the
nucleus of eukaryotic cells \cite{bronstein}.

\noindent
Explosions of solitons in a Kerr lens mode-locked Ti:sapphire laser were found experimentally
more than ten years ago \cite{cundiff2002}. However, very recently,
new evidence of exploding dissipative solitons has emerged from
an experiment in an all-normal-dispersion Yb-doped mode-locked fiber laser \cite{broderick}.
These pulses exhibit spatiotemporal chaos, thus, a feature of the experimentally observed explosions
is that they are similar but not identical to each other, in fact,
the times between explosions appear to be randomly distributed.
Although the real system is not continuous \cite{cundiff2002}, this dynamical
behavior was predicted theoretically in a continuous model, namely, the one-dimensional
complex cubic-quintic Ginzburg-Landau equation (CQGLE) \cite{soto2000}. There, these pulse solutions
present an unstable time evolution, but nevertheless, they remain confined in space. Large-amplitude
explosions might be considered as {\it extreme events}. Using this prototype equation (applicable
 near the weakly hysteretic onset of an oscillatory instability to traveling or standing waves \cite{bln})
 and by varying a single parameter, it has been found a transition from stationary pulses to exploding dissipative solitons (DSs)
via pulses that oscillate with one and two frequencies. This analogue of the Ruelle-Takens route for
spatially localized solutions indicates the chaotic nature of explosions \cite{orazio1,orazio2}. In addition, it
has been shown that the appearance of explosions has signatures of intermittency \cite{jaime1}.

\noindent
In the context of nonlinear optics, using an extended CQGLE in two spatial dimensions
\begin{eqnarray}
{\rm i} \, A_z &+& \frac{D}{2} \, \nabla^2_\perp A + |A|^2 A + \nu |A|^4 A
\nonumber\\
&=& {\rm i} \, \delta A + {\rm i} \, \varepsilon |A|^2 A + {\rm i} \, \beta \, \nabla^2_\perp A +
{\rm i} \, \mu |A|^4 A,
\label{GL1}
\end{eqnarray}

\begin{figure}[t]
\includegraphics[width=2.9in]{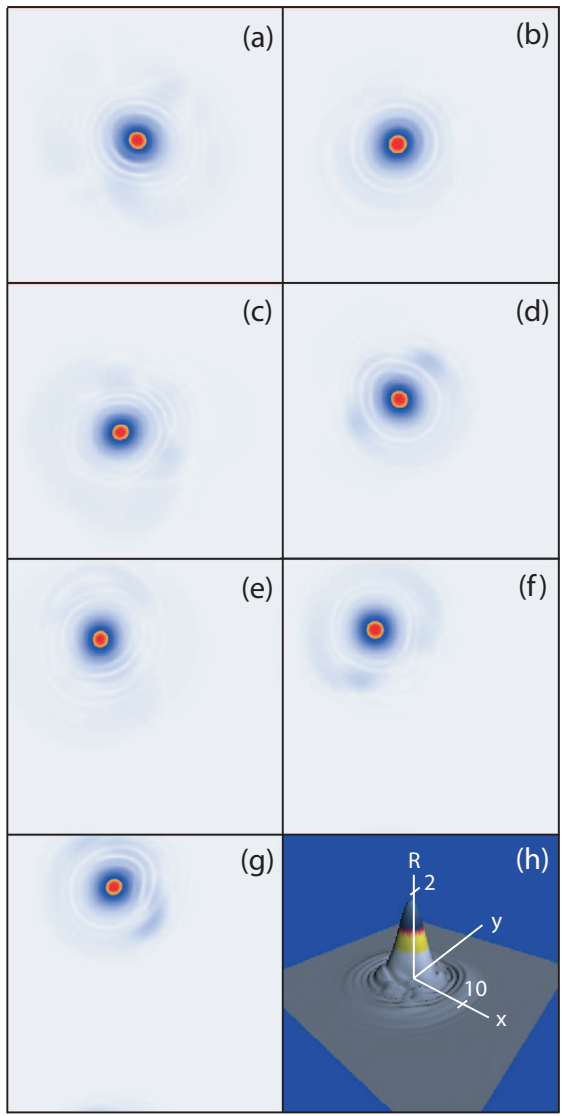}
 \caption{(a)--(g) 2-d Snapshots of the random time evolution of the modulus $|A(x,y;t)|$ of an
 asymmetric exploding soliton. $\mu = -0.1$, $\beta_r = 1$, $\beta_i = 0.8$, $\gamma_r = -0.1$,
$\gamma_i = - 0.6$, $D_r = 0.125$, $D_i = 0.5$. (a) $t = 0$; (b) $t = 200$;
 (c) $t = 400$; (d) $t = 600$; (e) $t = 800$; (f) $t = 1000$;
 (g) $t = 1200$; (h) 3-d snapshot of the DS between explosions. Maximum of
 $|A(x,y,t)|$ is around 2. In the plane $(x,y)$ the DS is localized inside
 a circle of radius 10.} \label{Bild1}
\end{figure}

\noindent it was shown that pulsating beams might evolve to exploding
DSs \cite{Devine}. Here $A(x,y,z)$ is the
normalized envelope of the electrical field, $\nabla^2_\perp$ stands for
the two-dimensional transverse Laplacian, $z$ is the propagation distance,
$D$ is the diffraction coefficient,
$\delta$ represents losses, and $\varepsilon$ is the nonlinear gain coefficient,
related to the pumping power.
The coefficient $\mu$ stands for the saturation of the nonlinear gain, while $\nu$
stands for the saturation of the Kerr
nonlinearity. Spectral filtering is characterized by the parameter $\beta$.

\noindent Using the transformation $z \rightarrow t$ and renaming the parameters as $\mu = \delta$,
$\beta_r = \varepsilon$, $\beta_i = 1$, $\gamma_r = \mu$, $\gamma_i = \nu$, $D_r = \beta$,
and $D_i = D/2$, the above equation can be transformed into the `hydrodynamical' form of the two-dimensional (2-d) cubic-quintic
complex Ginzburg-Landau equation
 \begin{eqnarray}
\partial_t A &=& \mu A + (\beta_r + {\rm i} \beta_i) |A|^2 A + (\gamma_r + {\rm i} \gamma_i) |A|^4 A \nonumber\\
&& + (D_r + {\rm i} D_i) \nabla^2 A,
\label{GL2}
\end{eqnarray}
\noindent where $A(x,y;t)$ is the envelope of the linearly unstable modes
at the onset of a subcritical instability, $\beta_r$ is postive and $\gamma_r$
negative in order to guarantee that the equation saturates to quintic order. As a
function of the bifurcation parameter $\mu$ and the nonlinear dispersion 
$\beta_i$, localized solutions have been studied in the
above equation \cite{orazio3}. Apart from stationary pulses, two types of exploding
dissipative solitons have been found. The first type, the azimuthally
symmetric exploding soliton, preserves the azimuthal symmetry even during
explosions and its center never moves. The second type,
the azimuthally asymmetric exploding soliton,
can suffer a shift in the position of its center during explosions.
In Fig.~\ref{Bild1} we show snapshots of the random
evolution of an azimuthally asymmetric exploding soliton between times involving around 20 explosions.
Each explosion (either symmetric or antisymmetric) takes around 1 or 2 units of time,
while the time between explosions lies in the range 20--30 units of time.

\begin{figure}[t]
\includegraphics[width=3in]{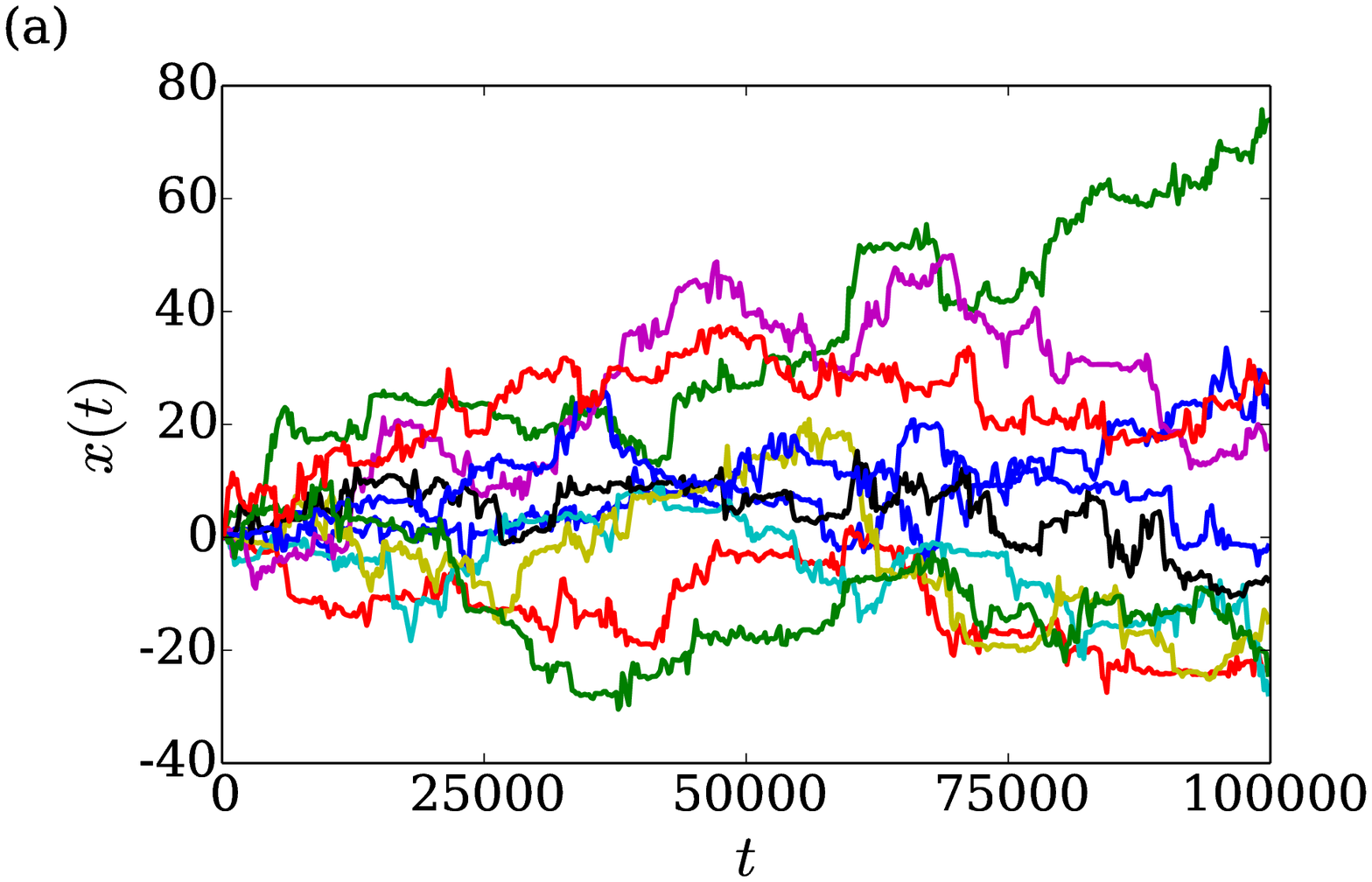}

\includegraphics[width=3in]{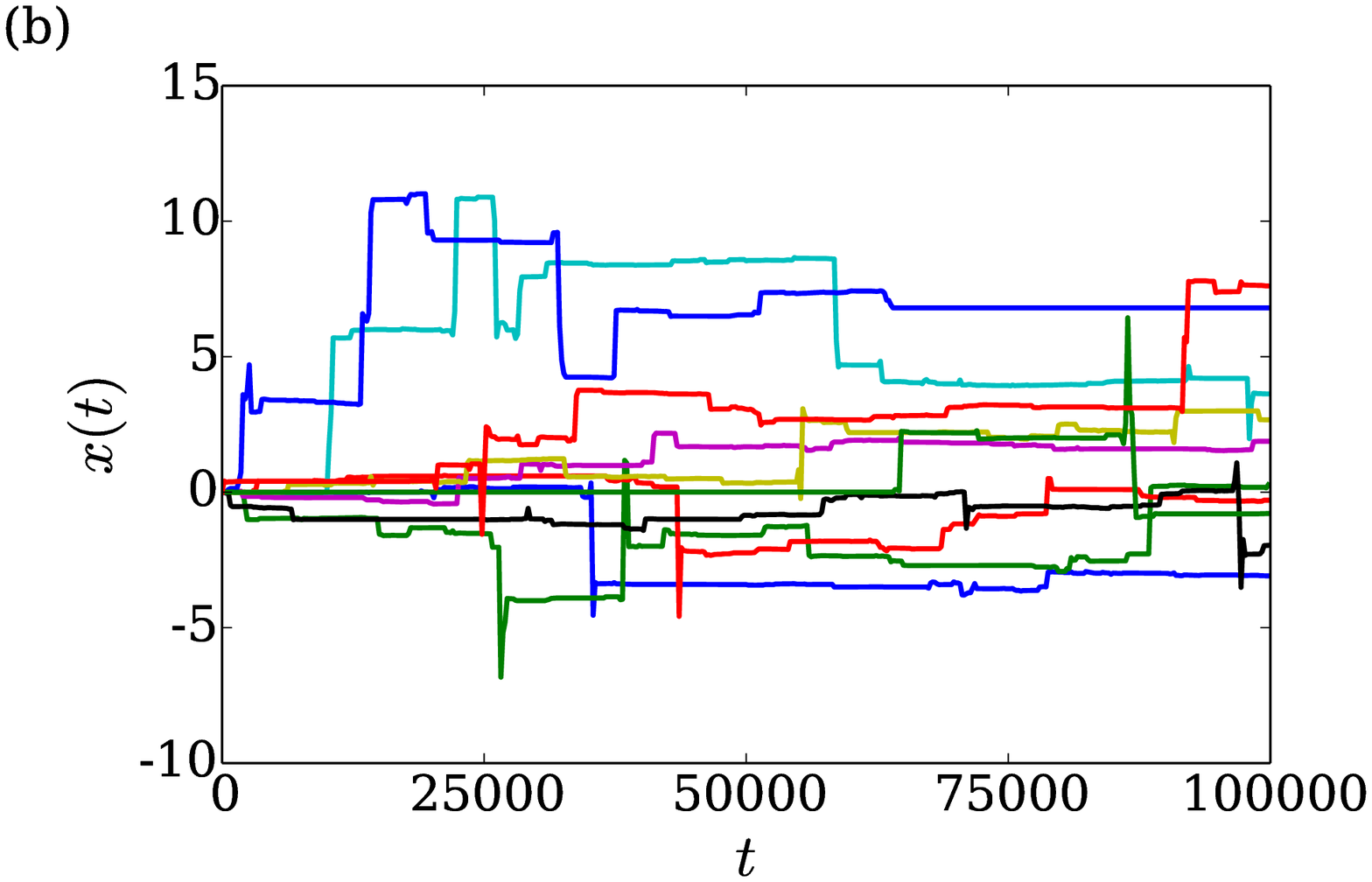}
\caption{Comparison of two different diffusive behaviors of the $x$-coordinate of the center of
mass of the soliton. For each value of $\mu$, the results of ten independent
realizations with slightly different initial conditions were superimposed.
(a) For $\mu=-0.30$. (b) For $\mu=-0.35$.
}
\label{fig:comparison}
\end{figure}

\noindent In this Letter we study the diffusive motion induced by
symmetric and asymmetric explosions. We  report anomalous
diffusion induced by intermittency between long sequences of
symmetric and asymmetric explosions.

\noindent
Eq.~(\ref{GL2}) was integrated from a localized initial condition, using a split-step
Fourier method, and the following parameters, which we kept fixed: $\beta_r = 1$,
$\beta_i = 0.8$, $\gamma_r = -0.1$, $\gamma_i = -0.6$, $D_r = 0.125$, and $D_i = 0.5$
corresponding to an anomalous dispersion regime in optics. The
parameter $\mu$ was varied from $-0.5$ to $-0.05$. The size of the domain was
chosen large enough so the amplitude was practically zero around the soliton. 
Simulations were carried out using a $256 \times 256$
spatial grid of size $dx \times dy = 0.2 \times 0.2$. The time discretization was $dt = 0.005$ and
runs typically involved $2 \times 10^7$ iterations so the total time was $T = 10^5$, and several
thousands of explosions were registered in each run.
The soliton persisted and remained localized at all times.

\begin{figure}[t]
\includegraphics[width=3in]{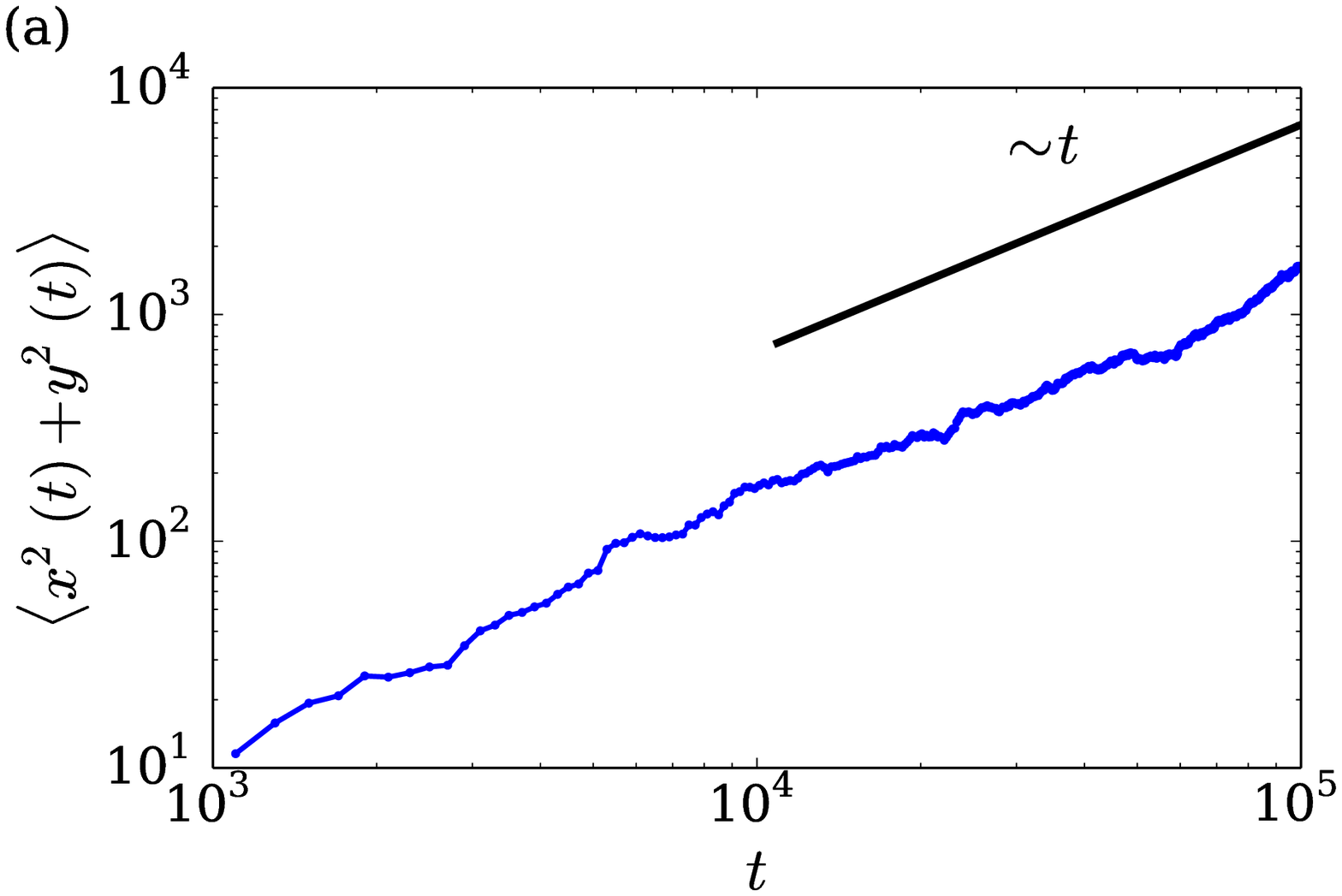}

\includegraphics[width=3in]{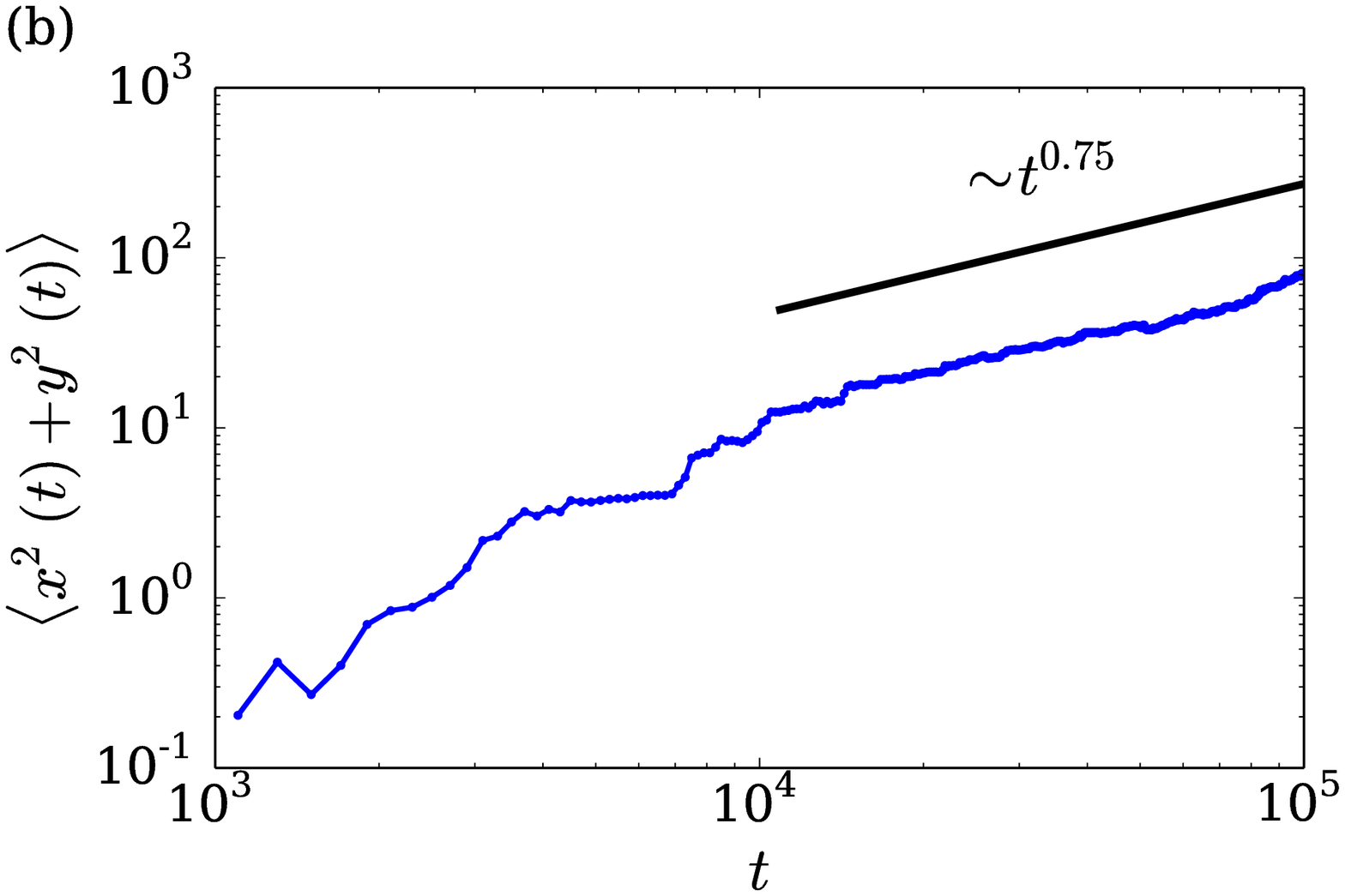}
\caption{Ensemble-averaged MSD defined by Eq.~(\ref{ensmsd}), evaluated for the same two cases
as in Fig.~\ref{fig:comparison}.
$N=128$ slightly different initial conditions were used in each case.
(a) For $\mu=-0.30$. (b) For $\mu=-0.35$.
}
\label{fig:comparison2}
\end{figure}

\noindent We found that there are two regimes of diffusive motion with qualitative differences.
On the one hand, there is a regime (for values of $\mu$ between $-0.32$ and zero) where most explosions
suffered by the soliton are asymmetric and lead to shifts.
On the other hand, there is another regime (for values of $\mu$ between $-0.37$ and $-0.32$, approximately), where the soliton suffers
both sequences of symmetric and asymmetric explosions \cite{orazio3}.
No sharp boundary between the two regimes has been found so far,
a rigorous analysis would require simulation times $T$ several orders of magnitude larger.

\noindent
As it was mentioned above, explosions exist in the one-dimensional complex CQGLE \cite{soto2000}. However, their
typical behavior is strongly asymmetric, being first reported in \cite{AS04}. Thus, this mechanism does not lead to
the anomalies present in 2-d, and the diffusion is normal.

\noindent
To quantify the differences between these two regimes we use the vector `center of mass' of the soliton:
$\vect{r}(t)=(x(t), y(t))$ for the density $|A(x,y;t)|^2$.
In Fig.~\ref{fig:comparison} we plot the $x$-coordinates of the centers of
mass of the dissipative soliton for the
two regimes, using a set of ten independent realizations for each choice of parameters.
Similar pictures can be obtained for the $y$-coordinate.
Each curve corresponds to the time evolution of an initial condition specified by a Gaussian
pulse plus spatial noise to generate the different realizations.
Fig.~\ref{fig:comparison}(a) shows for $\mu=-0.30$ an instance of the regime with only
asymmetric explosions in random directions.
Fig.~\ref{fig:comparison}(b) shows for $\mu=-0.35$ an instance of the regime with a
combination of symmetric and asymmetric explosions:
sequences of strictly symmetric explosions (not inducing spatial shifts) can be quite long, inducing
long periods of time where the coordinates of the soliton do not change.
In both regimes one can see how slight differences in initial conditions get magnified
over time as a result of explosions.

\begin{figure}[t]
\includegraphics[width=3in]{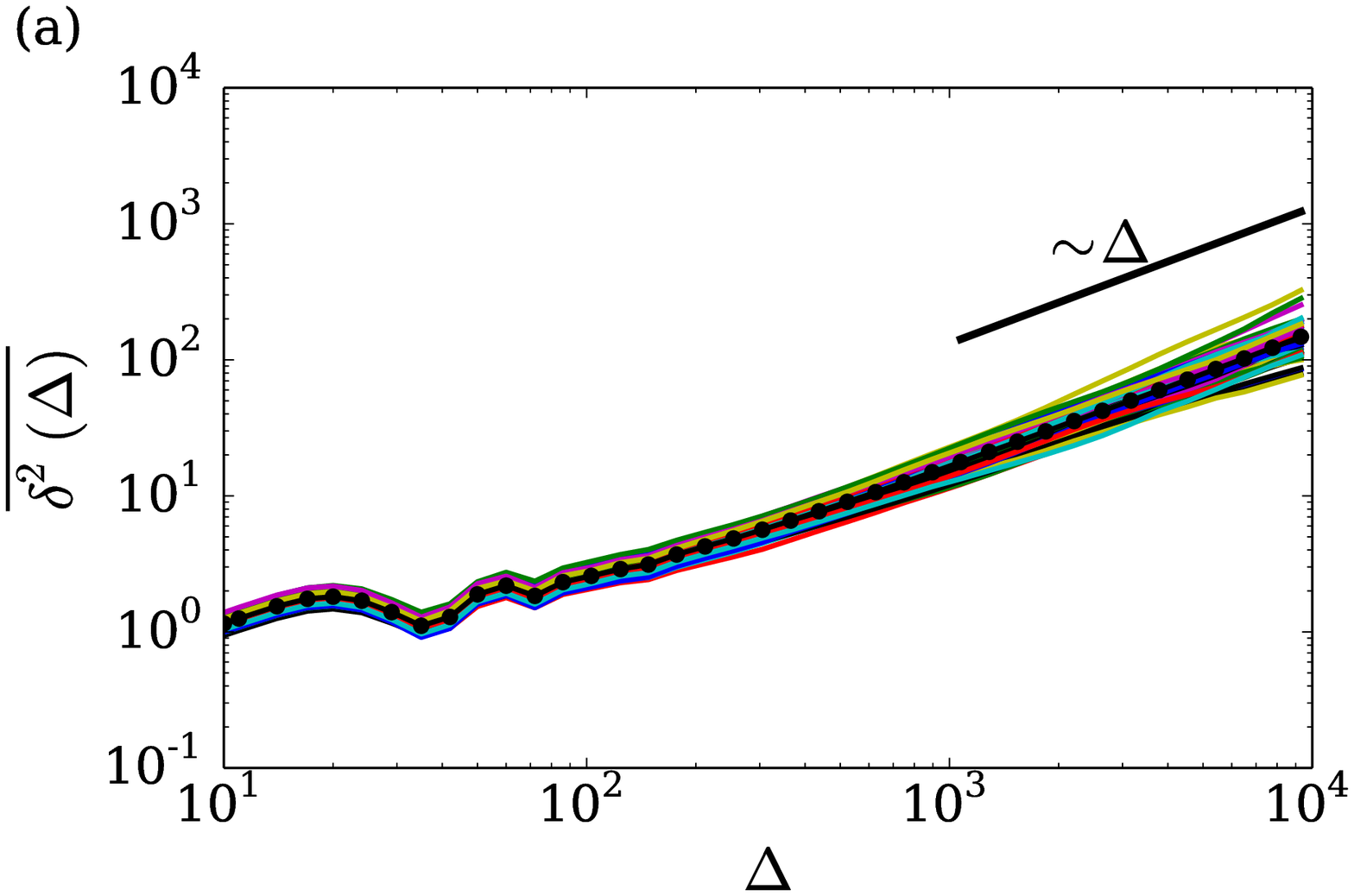}

\includegraphics[width=3in]{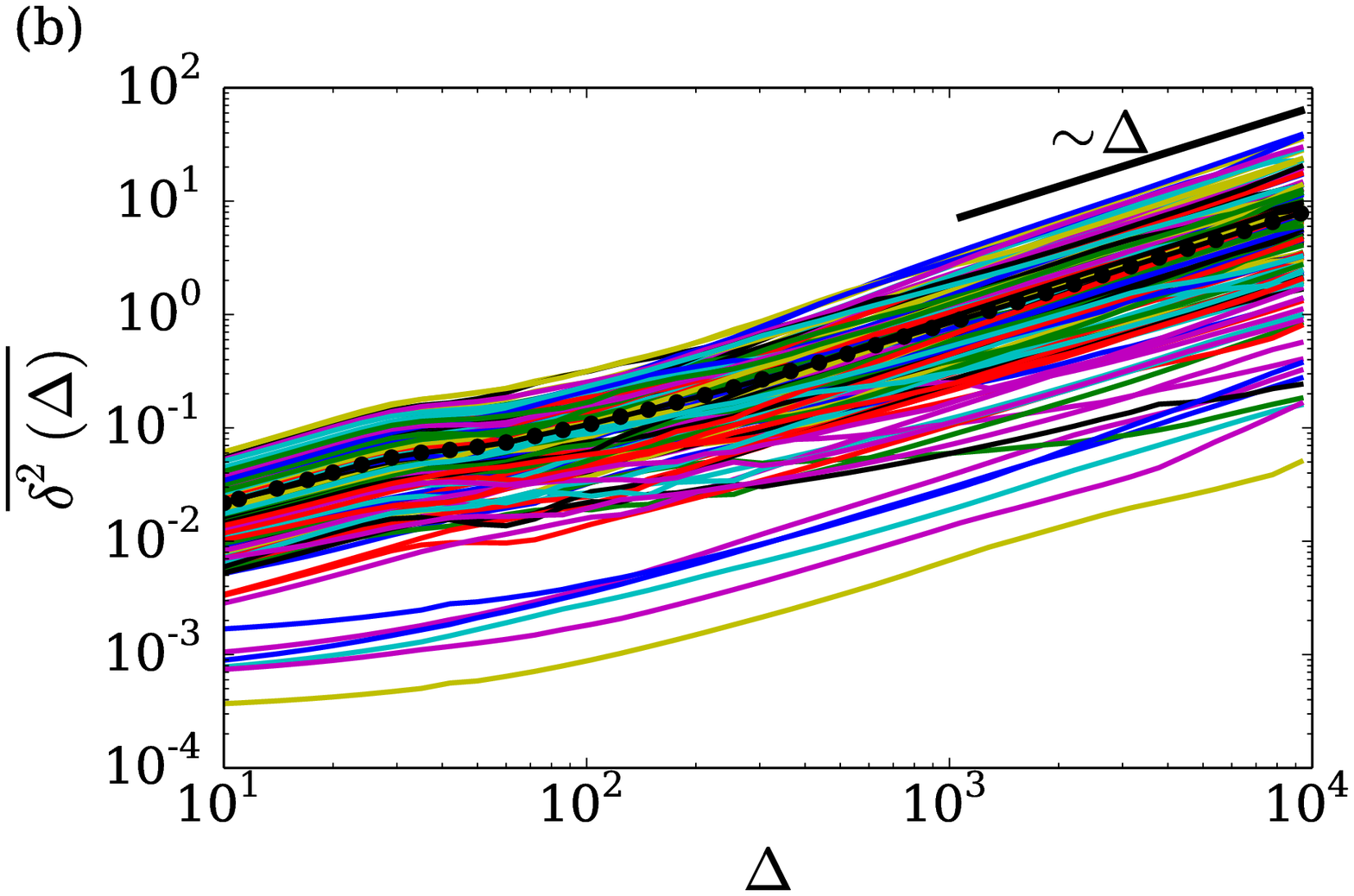}
\caption{Time-averaged MSD defined by Eq.~(\ref{tamsd}), evaluated for the same two cases
as in Fig.~\ref{fig:comparison}.
$N=128$ slightly different initial conditions were used in each case.
(a) For $\mu=-0.30$. (b) For $\mu=-0.35$.
}
\label{fig:comparison3}
\end{figure}

\noindent The regime presented in Fig.~\ref{fig:comparison}(b) with solitons suffering both
symmetric and asymmetric explosions seems to show some features of \emph{subdiffusion}.
Anomalous diffusion, that includes subdiffusion, superdiffusion and other exotic phenomena,
has been analyzed using a sophisticate mathematical framework developed in the last decades
\cite{pccp}. In this work we will assess whether the motion of the soliton qualifies as anomalous,
more specifically, whether it can be modeled by a subdiffusive continuous-time random walk (CTRW) \cite{MW65,MK00,AR13}.

\noindent The basic quantity that is used to characterize diffusive motion is
the mean squared displacement (MSD).
Now, in the framework of anomalous diffusion, one has to distinguish between
two different definitions of MSD.

\noindent The \emph{ensemble-averaged} MSD is given by:
\begin{equation}
\langle || \vect{r}(t) ||^2 \rangle \mydef \lim_{N \rightarrow \infty} \frac{1}{N} \sum_{i=1}^N || \vect{r}_i(t) ||^2
\label{ensmsd}
\end{equation}
assuming all initial conditions verify $\vect{r}_i(0)=0$.
In Fig.~\ref{fig:comparison2} we plot the average from an ensemble of $N=128$ independent realizations
for the same representative cases ($\mu = -0.30$ and $\mu = -0.35$) presented in Fig.~\ref{fig:comparison}.

\begin{figure}[t]
\includegraphics[width=3in]{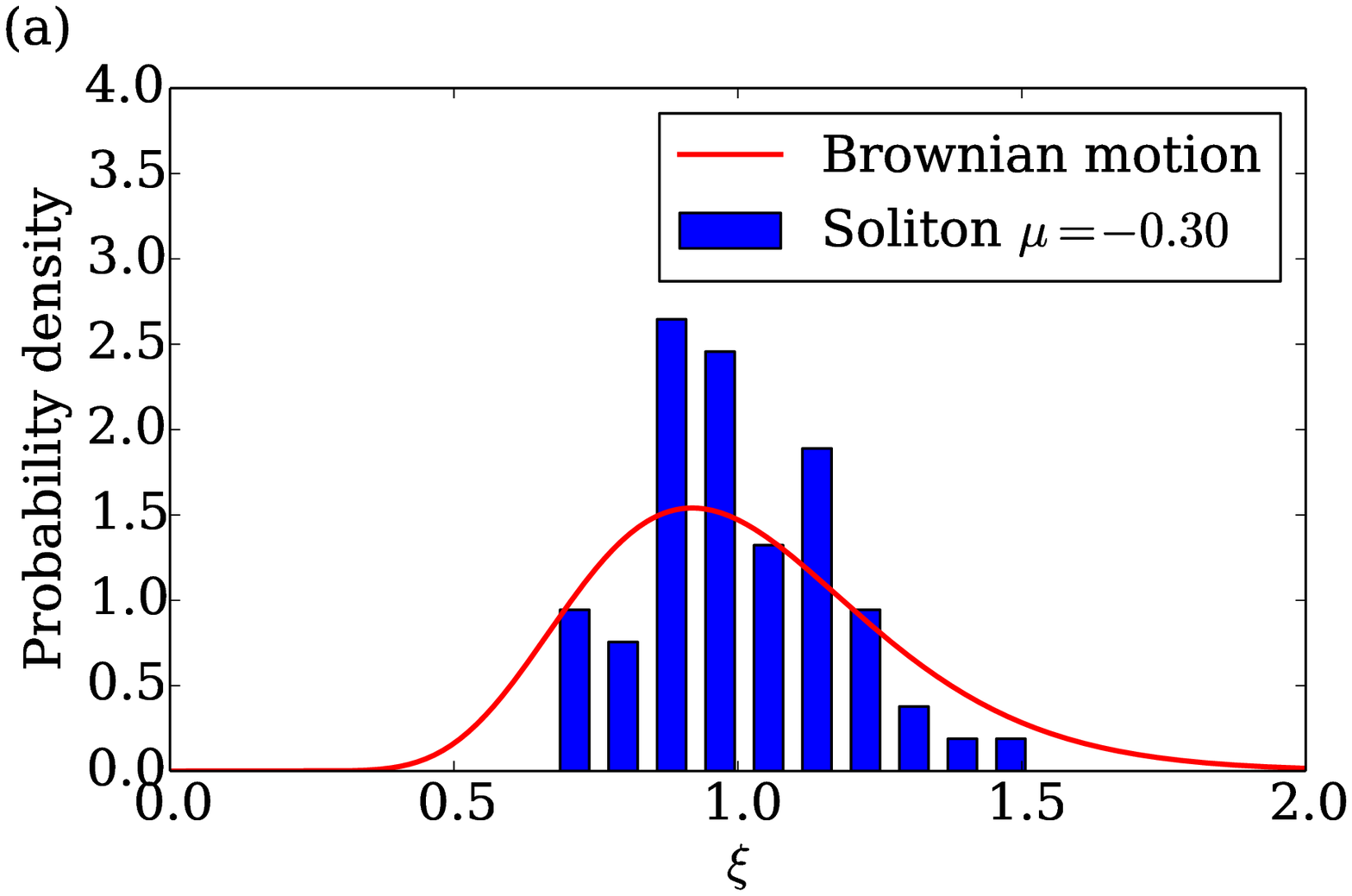}

\includegraphics[width=3in]{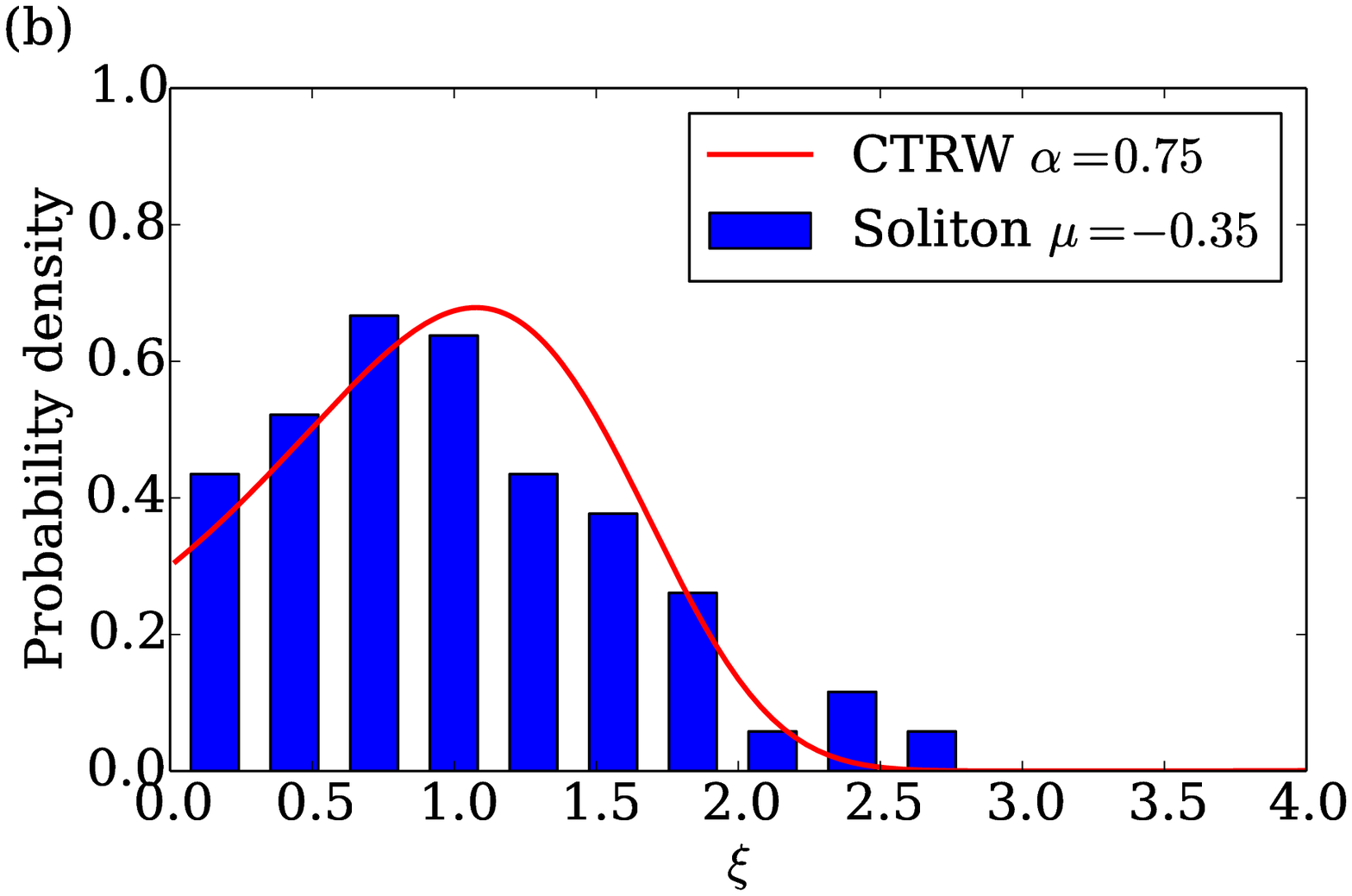}
\caption{Statistics of the normalized time-averaged MSD defined in Eq.~(\ref{defxi})
(with $\Delta=10^4, T=10^5$) extracted from the soliton behavior.
(a) For $\mu=-0.30$, the histogram of $\xi$ is compared with $\phi(\xi)$ associated to Brownian motion.
(b) For $\mu=-0.35$, the histogram of $\xi$ (using $\Delta=10, T=10^5$) is compared with the distribution defined in Eq.~(\ref{phi_xi})
associated to a subdiffusive 2-d CTRW,
using $\alpha=0.75$ extracted from the ensemble-averaged MSD (see Fig.~\ref{fig:comparison2}b).
}
\label{fig:phi_xi}
\end{figure}

%

\noindent The ensemble-averaged MSD seems to increase for long times as
$$
\langle || \vect{r}(t) ||^2 \rangle \sim t^\alpha ,
$$
but with different exponents in the two cases.
For $\mu=-0.30$ the exponent seems to be 1, which is consistent with the
shape of the trajectories and the basic assumptions of normal diffusion: uncorrelated
shifts (asymmetric explosions) of finite size and finite characteristic time.
For $\mu=-0.350$, however, there are indications of subdiffusion: $\alpha<1$.

\noindent The \emph{time-averaged} MSD of an individual trajectory reads:
\begin{equation}
\overline{\delta^2(\Delta,T)} \mydef \frac{1}{T-\Delta} \int_0^{T-\Delta} || \vect{r}(t+\Delta)-\vect{r}(t) ||^2 \ dt ~,
\label{tamsd}
\end{equation}
with a lag-time $\Delta<T$ (in our analysis we use $0<\Delta<T/10$).
Our results for the same two cases, see Fig.~\ref{fig:comparison3}, indicate that for long $T$ we get
$$
\overline{\delta^2(\Delta,T)} \approx D_{ta}\ \Delta,
$$
where $D_{ta}$ is a generalized diffusion coefficient.

\noindent In the normal regime $\mu=-0.30$, the estimated $D_{ta}$ from individual realizations
take similar values.
But in the subdiffusive regime $\mu=-0.35$,
the estimated $D_{ta}$ vary substantially from trajectory to trajectory and show a wide distribution.
In the log-log plot of Fig.~\ref{fig:comparison3}(b), the variability of $D_{ta}$ appears
as a variable position coefficient of the straight lines
corresponding to individual realizations.
(Other examples of this phenomena and rigorous analyses can be found in Refs. \cite{GC06} and \cite{pccp}.)

\noindent
Comparing Fig.~\ref{fig:comparison2}(b) with Fig.~\ref{fig:comparison3}(b) it is clear that
for $\mu=-0.35$, where the ensemble average shows subdiffusion,
and for finite $T$ and $\Delta<T$, one has
$ \overline{\delta^2(\Delta,T)} \ne \langle || \vect{r}(\Delta) ||^2 \rangle ~,$
and there are no indications that the equality might hold in the limit of long simulation time
i.e. the exponent $\alpha$ of the ensemble-averaged MSD does not seem to approach one.
The last inequality indicates the phenomenon of `weak ergodicity breaking' since
the ensemble average cannot be
reproduced from a time average of a very long realization \cite{pccp}.
A more systematic way to analyze the scatter of the
individual realizations is to look at the distributions of the normalized time-averaged MSD:
\begin{equation}
\xi \mydef \frac{\overline{\delta^2(\Delta,T)}}{\langle \overline{\delta^2(\Delta,T)} \rangle} ~.
\label{defxi}
\end{equation}
In case that all trajectories show the same exponent $\alpha$,
$\xi$ can also be understood as the normalized diffusion coefficient $D_{ta}/\langle D_{ta} \rangle$.
In Fig.~\ref{fig:phi_xi} we show the histograms for $\xi$
estimated from individual trajectories in the two regimes.

\noindent
For the normal regime, the histogram can be compared with
the Laplace inversion of the generalized Gamma distribution \cite{AG12},
which approximates the exact behavior of Brownian motion.

\noindent
For the subdiffusive regime, the motion of the soliton
(ignoring the small time taken by each explosion)
can be compared with a 2-d CTRW,
defined with a waiting-time distribution $\psi(t_w) \sim 1/t_w^{1+\alpha}$
and a jump-size distribution $\lambda(\vect{r})\sim \delta(||\vect{r}||-1)$.
The analysis of the times between asymmetric explosions of the soliton
shows a probability distribution that is consistent with a power-law,
and the parameter $\alpha=0.75$ (extracted from the ensemble-averaged MSD in Fig.~\ref{fig:comparison2}b).
This value of $\alpha$ implies an infinite mean waiting-time,
a key feature of subdiffusive CTRW (see Appendix).

\noindent
Based on this simplification, one can now compare the statistics of $\xi$ extracted from
the numerical simulation of the soliton with the theoretical results for the CTRW.
For this last model, $\xi$ depends on the random number $n$ of
jumps which occur up to time $T$. Obviously, $n$ does not depend on the spatial
dimension of the jumps but is only influenced by the waiting times.
The random variable $\xi$ is then equal in distribution with the
random variable $n/\langle n \rangle$, and for large $T$ and small $\Delta$,
it can be approximated as:
\begin{equation}
\phi(\xi) \simeq \frac{\Gamma^{1/\alpha}(1+\alpha)}{\alpha \xi^{1+1/\alpha}}\ \ell_\alpha \left( \frac{\Gamma^{1/\alpha}(1+\alpha)}{\xi^{1/\alpha}} \right) ~,
\label{phi_xi}
\end{equation}
where $\ell_\alpha(\cdot)$ is the one-sided L\'evy stable distribution,
whose Laplace transform is $\exp(-s^\alpha)$ \cite{HBMB08}.
As depicted in Fig.~\ref{fig:phi_xi}(b)
the theoretical curve shows similar features as the soliton data.
The variance of $\xi$ is known as the \emph{ergodicity breaking} parameter,
which for the Mittag-Leffler distribution reads:
$$
EB_\mathrm{CTRW}(\Delta) \mydef \lim_{T \rightarrow \infty} \langle \xi(\Delta,T)^2 \rangle -1 =
\frac{2\Gamma^2(1+\alpha)}{\Gamma(1+2\alpha)}-1 ~.
$$
\noindent
For $\alpha=0.75$ this last equation gives $EB_\mathrm{CTRW}=0.27$ that is smaller than
the estimate from the subdiffusive explosive soliton, $EB=0.39$,
which can be attributed to the finite length, $T=10^5$, of the simulated trajectories.

\noindent
We have not observed a persistence in the direction of the shifts of the soliton.
Long sequences of shifts with the same direction could in principle balance the effect of long waiting-times
and even lead to superdiffusion \cite{SSRS15}.

\noindent
In summary, we have shown anomalous diffusion and weak ergodicity breaking
of dissipative localized structures
in a `simple' and deterministic prototype model: the cubic-quintic
complex Ginzburg-Landau equation in two spatial dimensions. The ensemble-averaged mean squared
displacement gives an exponent lower than 1, corresponding to subdiffusion;
while the time-averaged mean squared displacement of an individual trajectory
grows linearly with time-lag.
This anomalous regime shows many similarities with a subdiffusive continuous-time random walk.

\begin{acknowledgments}
J.C. and O.D. are thankful for the support of FONDECYT (Chile, Grants 1140143 and 1140139),
and the Research Office, Universidad de los Andes, Chile.
\end{acknowledgments}


\appendix

\section{Waiting-time distributions between asymmetric explosions}

\begin{figure}[t]
\begin{center}

\includegraphics[width=3in]{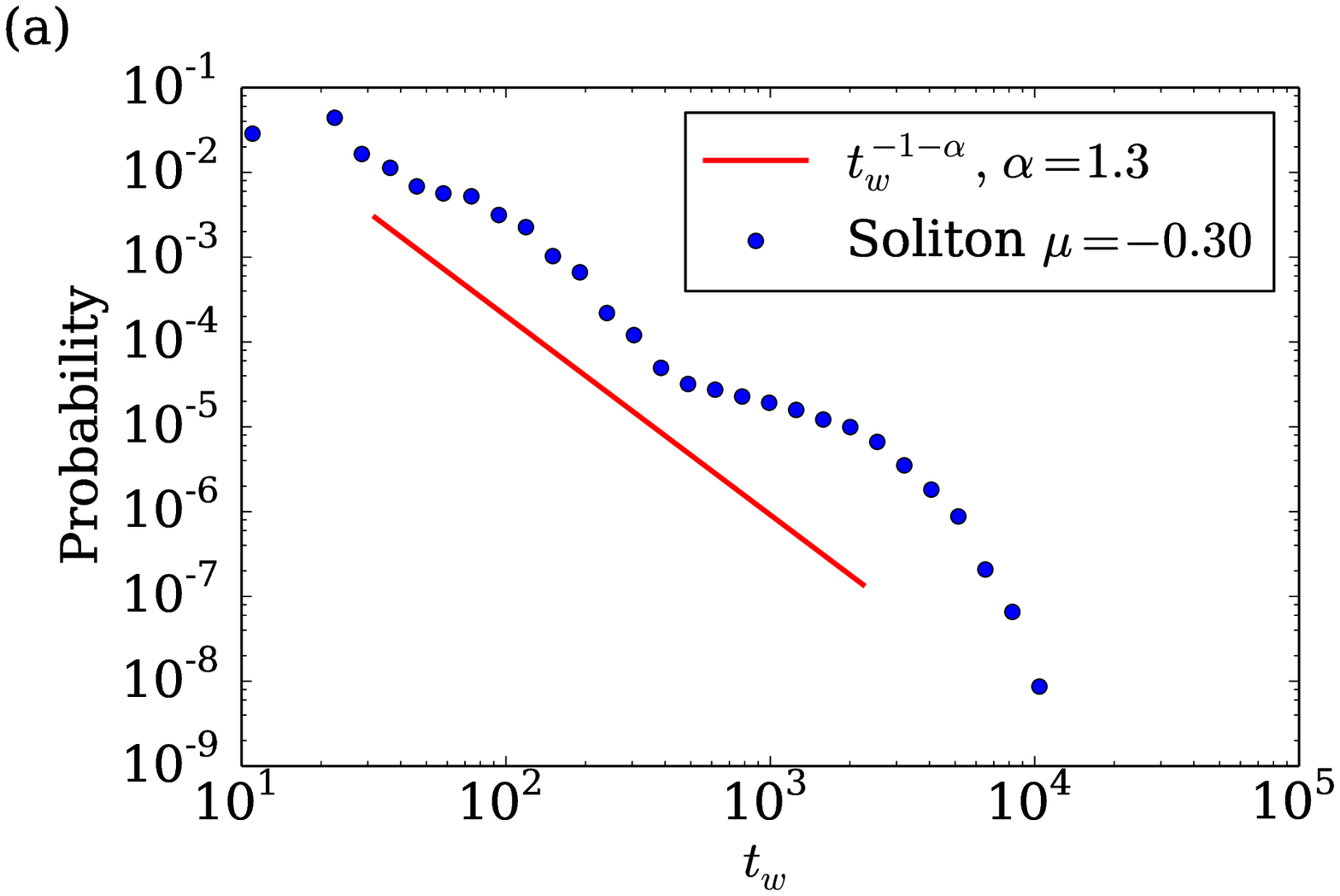}

\includegraphics[width=3in]{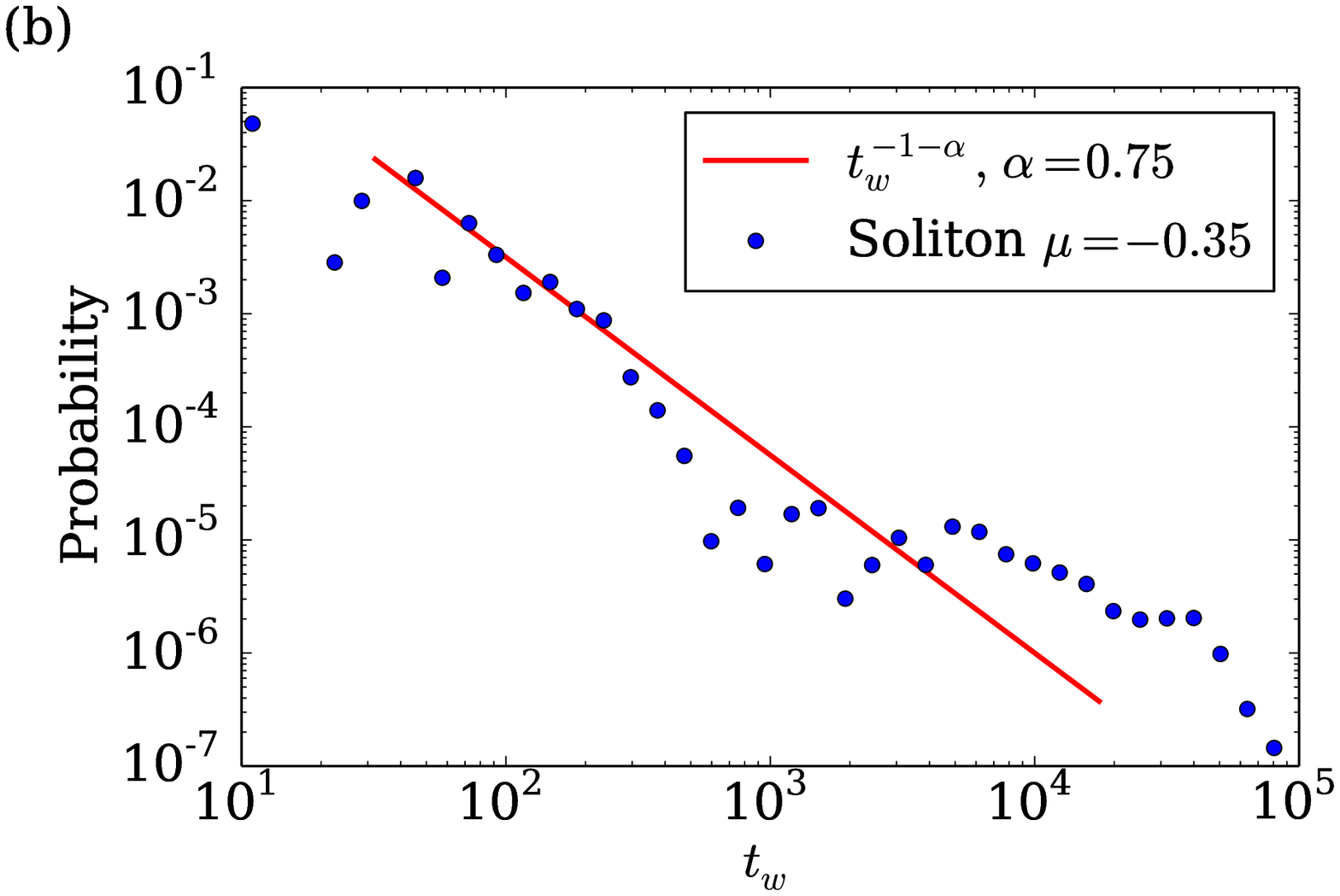}
\end{center}
\end{figure}

The apparently random motion of the soliton induced by the intermittent asymmetric explosions can be compared
with a 2-d continuous-time random walk (CTRW), that is defined by the jump-size distribution $\lambda(\vect{r})$
and the waiting-time distribution $\psi(t_w)$. Different choices for these distributions lead to a wealth of behaviors.

For the dissipative soliton, the spatial shifts are always comparable to its width,
so we can use $\lambda(\vect{r})\sim \delta(||\vect{r}||-1)$ for the analysis.
The statistics of the waiting-times can be captured using a power-law $\psi(t_w) \sim 1/t_w^{1+\alpha}$.

The figures show the waiting-time probability distributions $\psi(t_w)$ for the asymmetric explosions.
(a) For $\mu=-0.30$. (b) For $\mu=-0.35$.
For both cases
the corresponding waiting time distribution
decays roughly with a power law $\psi(t_w) \sim 1/t_w^{1+\alpha}$, with $\alpha=1.3$ (a) and $\alpha=0.75$ (b)
implying a finite and an infinite mean waiting time, respectively.
While the finite mean waiting time leads to normal diffusion (see Fig.~3a in the Letter),
the infinite mean waiting time is the key feature of a CTRW with subdiffusive character,
as observed in Fig. 3(b) in the Letter. 

\end{document}